\documentclass[12pt,preprint]{aastex}

\shorttitle{{\it Spitzer} IRS study of MF16 ULX}
\shortauthors{Dudik et al.}
\begin{document}

\title{{\it Spitzer} Observations of MF 16 Nebula and the associated Ultraluminous X-ray Source}

\author{C. T. Berghea and R. P. Dudik} 
\affil{United States Naval Observatory, Washington, DC 20392}
\email{ciprian.berghea@usno.navy.mil}
\email{rpdudik@usno.navy.mil}

\begin{abstract}
We present {\it Spitzer} Infrared Spectrograph (IRS) observations of the ultra-luminous X-ray source (ULX) NGC~6946 X-1 and its associated nebula MF~16. This ULX has very similar properties to the famous Holmberg~II ULX, the first ULX to show a prominent infrared [O~IV] emission line comparable to those found in AGN. This paper attempts to constrain the ULX Spectral Energy Distribution (SED) given the optical/UV photometric fluxes and high-resolution X-ray observations. Specifically, {\it Chandra} X-ray data and published {\it Hubble} optical/UV data are extrapolated to produce a model for the full optical to X-ray SED. The photoionization modeling of the IR lines and ratios is then used to test different accretion spectral models. While either an irradiated disk model or an O-supergiant plus accretion disk model fit the data very well, we prefer the latter because it fits the nebular parameters slightly better. In this second case the accretion disk alone dominates the extreme-UV and X-ray emission, while an O-supergiant is responsible for most of the far-UV emission.

\end{abstract}


\section{INTRODUCTION}

The suggestion that ultraluminous X-ray sources (ULXs) in nearby galaxies may harbor intermediate-mass black holes  \citep[IMBH, e.g.][]{col99} has major implications on the studies of black hole formation and evolution. However, the evidence for this new type of black hole has been slow to accumulate.  Other explanations have been proposed to account for the high X-ray luminosities of ULXs, such as super-Eddington accretion  \citep{done06, kun07, king08, sor08, ber08, abol11} or beaming \citep{kor02, king08}. Some galactic binaries are known to show super-Eddington luminosities, such as GRS~1915+105 \citep{fen04} and V4641~Sgr \citep{rev02}.  However conclusive evidence on the true nature of some or all known extra-galactic ULXs has remained elusive. 

One of the most interesting developments in ULX history is the discovery in recent years of large ionized bubble nebulae around some of the most famous ULXs \citep{pak02}.  These are much larger and more luminous than regular supernova remnants. In addition, high-ionization lines have been detected, such as He~II $\lambda$4686 \citep[I.P.~$=$~54.4~eV; e.g.][]{abol07, moon11}. \citet{kaa09} detected [Ne~V] $\lambda$3426 from the nebula associated with the ULX in NGC~5408 (I.P.~$=$~97.1~eV).  The detection of these lines suggests photoionization of the surrounding gas by the hard radiation from the accreting ULX, similar to what happens with active galactic nuclei (AGN).

It is also remarkable that there may exist such an object in our own galaxy.  The first recognized microquasar, the famous SS~433 \citep{sanduleak} is possibly the ``closest'' relative to the nebulae-surrounded ULXs.
It is surrounded by a 100~pc radio nebula, W~50 \citep{dub98} and is most likely super-Eddington  \citep[e.g.][]{fab04}. Indeed, SS~433 could be an example of both beaming {\it and} super-Eddington accretion, the combination of which could easily explain even the most luminous ULXs \citep{beg06, pou07}. It is seen almost edge-on and therefore most of its strong UV and X-ray radiation is likely not visible.  However, if such objects exist in nearby galaxies, those seen at small inclination angles should appear as bright ULXs.  Recently an object very similar to SS~433 was discovered in the galaxy NGC~7793 \citep{pak10, sor10}, showing radio lobes and hot spots, like a bona-fide quasar.  

\citet{vaz07} conducted the first infrared spectroscopy study of ULXs.
They obtained high-resolution {\it Spitzer} spectra of six ULXs in NGC~4485/90.
\citet{vaz07} looked for the mid-infrared [O~IV] high ionization line, which has an equivalent ionization potential to He~II.  However, apart from a possible weak detection in one object in their sample (ULX-1, which is probably a supernova remnant), [O~IV] was not found in the NGC~4485/90 ULXs.

The first reliable detection of [O~IV] associated with a ULX was the famous Holmberg~II ULX \citep{ber10a, ber10b}.  It is also one of the most interesting examples of an ionized nebula associated with a ULX \citep{pak02,leh05, abol07}. He~II is detected in a small nebular region around the ULX, the ``Heel'' of the ``Foot Nebula'' \citep{pak02, kaa04}. An OB star is the likely counterpart.  The morphology of He~II emission and other optical lines is consistent with the photoionization of the nebula by the ULX  \citep{pak02}. \citet{kaa04} confirmed this using the Hubble Space Telescope (HST). \citet{mill05} detected extended radio emission from a $\sim$50~pc diameter region coincident with the position of the ULX.  These authors conclude that the flux and morphology of the radio emission is inconsistent with emission from a SNR or an H~II region, and is probably associated with the ULX activity. Using new observations, \citet{cseh12} suggest that the radio emission is probably caused by a weakly collimated outflow, rather than a jet.

The ULX which we investigate in this paper (which we will call NGC 6946~X-1 from now on) was first identified as such by \citet{rob03}, who showed that the X-ray variability of this source indicates a binary origin.  MF~16 has been known as a very luminous SNR \citet{mf97} associated with a ROSAT X-ray source and a radio source \citep{dyk94}. \citet{abol08} used optical spectroscopy to conclude that the ULX is most likely responsible for the high ionization lines detected in the MF~16 nebula.  They suggest that the ULX is probably bright in UV/EUV and not only in X-rays (predicted absorbed flux at 1000~$\AA$: 10$^{-17}$ erg cm$^{-2}$ s$^{-1}$ $\AA$ $^{-1}$). This was actually verified by \citet{kaa10}, who used the ACS camera on the Hubble Space Telescope to detect far-UV emission from the ULX for the first time. The data does not allow the authors to conclude with certainty what the UV source is, but they suggest it is a combination of disk emission and contribution from an OB supergiant.

Recently a tentative quasi-periodic oscillation of $\sim$8.5~MHz was detected \citep{rao10} using XMM-Newton X-ray data which - if confirmed - makes the NGC 6946~X-1 the fourth ULX with such timing features (2 ULXs in M82 and NGC~5408~X-1).  NGC 6946~X-1 is very similar to the Holmberg~II ULX in many respects. \citet{ber10b} used Mappings~III photoionization and shock modeling to show that NGC 6946 X-1 and Holmberg~II ULX high-ionization line fluxes primarily result from photoionization rather than shocks (see also Section~4.1.).

In this paper we use IR {\it Spitzer} spectroscopic observations to constrain the UV and soft X-ray emission of the NGC~   6946 X-1.  The IR observations, when combined with previously published optical, UV and X-ray data, suggest that the powerful UV and soft X-ray emission from NGC 6946 X-1 photoionizes the surrounding medium. In Section~2 we present the {\it Spitzer} IRS data analysis and {\it Chandra} X-ray spectral fitting results. In Section~3 we use our data and other published data to construct an SED of the ULX which is then used as photoionization source in CLOUDY modeling.  Finally, in Section~4 we use our results to compare the ionization properties of this famous ULX with those of standard AGN.

\section{DATA PROCESSING AND RESULTS}

\subsection{{\it Spitzer} IRS Observations}

The data presented here are archival {\it Spitzer} data, taken on 2009 February~21.
The IRS spectral map is described by four parallel pointings and 12 perpendicular at half-slit steps, for the Short High (SH, in the range 9.9-19.6~$\mu$m) module.  The Long High map (LH, in the range 18.7 $-$ 37.2~$\mu$m) has three parallel pointings and 7 perpendicular. 

The total exposure time was 1.4~ks and 1.3~ks for SH and LH observations, respectively.   The data were preprocessed by the IRS pipeline (version 18) at the {\it Spitzer} Science Center prior to download. Spectral maps were constructed using ``BCD-level'' processed data in conjunction with CUBISM (CUbe Builder for IRS Spectra Maps) version 1.5\footnote{Available at: http://ssc.spitzer.caltech.edu /archanaly/contributed/cubism/} \citep{smith07, ken03}. The absolute calibration error on fluxes obtained with CUBISM is $\sim$25\% \citep{smith04}.
We adopt a distance of 5.1~Mpc to NGC~6946 as in \citet{kaa10}(1$\arcsec$ is 25~pc).  The LH maps obtained with CUBISM have intrinsic sizes of 26.4$\arcsec$ by 17.6$\arcsec$ (660~pc by 440~pc), while the SH maps are 23$\arcsec$ by 13.8$\arcsec$ (575~pc by 345~pc). However, in order to avoid aperture effects, the spectra presented here were extracted from a like-sized aperture. Thus the SH and LH spectral data result from an equally sized square aperture of size 13.3$\arcsec$ (332.5~pc), which is composed of 9 LH pixels centered on the ULX position.

\begin{figure}
\epsscale{0.6}
\plotone{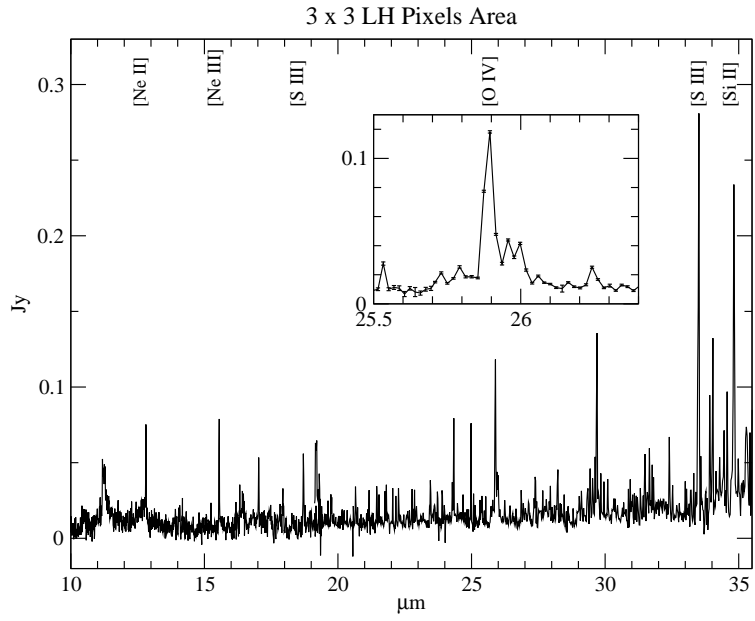}
\caption{
Full IRS spectra obtained from matched apertures with CUBISM from the aperture shown in Fig.~\ref{maps}} 
\label{spectra}
\end{figure}

The high-resolution spectra (10~$\mu$m$-$37~$\mu$m) are plotted in Figure~\ref{spectra}.  Since the spectra are extracted from matched apertures, the spectral lines in both the SH and LH modules are properly calibrated and the continuum is properly matched. We used SMART \citep[Spectroscopic Modeling Analysis and Reduction Tool, see][]{hig04} version 8.1.2 to measure line fluxes.  The fluxes of detected lines and upper limits for non-detections are presented in Table~\ref{table1}. We note that the [O~IV] line is detected at a signal-to-noise ratio $>$~12.
\citet{abol08} estimates an extinction of A$_V=~$1.54 mag in the direction of NGC 6946 X-1.  For the infrared lines, we estimate from \citet{draine} that the extinction is A$_{\lambda}<$~0.1, and therefore negligible. 

The SH and LH maps for the detected lines are presented in Figure~\ref{maps}.  They are shown together with two Hubble Space Telescope ({\it HST}) images of the MF~16 nebula in UV and H$\alpha$ emission.  A nearby (200~pc away) star-forming region and the associated HII region can be seen just outside the {\it Spitzer} aperture.  As can be seen in the {\it Spitzer} spectral maps, the low-ionization lines ([Ne~II], [S~III], and [Si~II]) are contaminated by the surrounding star-forming region. In contrast, the higher ionization lines [O~IV] and [Ne~III] are clearly concentrated at the NGC~6946 X-1 location, suggesting a source of high-energy photons.

\subsection{Archival Data}
\subsubsection{\bf X-ray spectrum:}
We use X-ray spectroscopy in conjunction with published optical/UV data to model the X-ray emission from the ULX and to extract the full optical to X-ray spectrum. 
We also use it to estimate the UV emission by extrapolating the model to lower energies. To model the X-ray spectrum we used the best available data from the longest ($\sim$58~ks) {\it Chandra} observation in the archive (Obs. ID 1043, taken on 2001 September 7).  The data were processed by \citet{ber08} (U45 in this paper) and the spectrum is plotted in Figure~\ref{sed}.

We note that \citet{frid08} find long-term X-ray flux variability of at most 30\% in NGC~6946 X-1 based on {\it Chandra}, XMM-Newton and ROSAT data.  However for completeness we have chosen the best data available to represent the source (e.g. {\it Chandra} data rather than ROSAT data), as well as one in which the luminosity is approximately average over the 10 years of observations (i.e. the 2001 data set).  This average X-ray spectrum and luminosity best represents the ULX in a state in which the IR-line fluxes can be modeled.  In addition, the equilibrium time-based recombination for the high ionization lines are approximately 3000 to 100 years for He~II and [O~IV] respectively$-$well above the variability time-scales of the X-ray source \citep{ber10b}.

\subsubsection{\bf HST UV and optical photometry:} 
We use the measurements obtained by \citet{kaa10} to constrain the optical and UV emission from the accreting source and its companion. The B, V, I and UV (F140LP filter) extinction-corrected fluxes are plotted as circles in Figure~\ref{sed}.

\section{CLOUDY MODELING}

\subsection{Fitting the X-ray Spectrum and Optical Data}
We used XSPEC version 12.4 to model the spectrum between 0.3 and 8.0~keV. 
The X-ray spectrum is the most important component of the total SED since it is likely the only observed emission that directly emanates from the ULX, and is not contaminated by the surrounding star-forming region.  For this reason we have taken many precautions in modeling the X-ray spectrum, however we fit the X-ray and optical/UV literature data simultaneously.  In this section we describe the models we chose to fit the X-ray spectrum and extrapolate to the unknown UV region not covered by literature data.  

Probably the most common and simplest model for ULX X-ray spectra is the two-component model with a multi-color accretion disk component \citep[MCD,][]{mit84} and a power-law for the hard tail. However, this model cannot be extrapolated to UV where the source of IR-lines originates \citep[e.g.]{done06}, and therefore we use the physically motivated Comptonization models described in detail below.

For all models, following \citet{ber08}, the Galactic absorption column density was fixed at 2.02$\times$10$^{21}$ cm$^{-2}$, and an additional local absorption (N$_H$) model was added. We assume that the metallicity is approximately Solar \citep[see][and references therein]{gar10}. 

{\bf DISKIR Model:}  The DISKIR model in XSPEC is a particularly useful Comptonization model to fit both optical and X-ray data, because it also includes emission from an irradiated disk \citep{gie09}. We found that the best fit model has a disk of radius 500~R$_{in}$ (the inner disk radius), and the irradiated flux is at 1.5\% of the total flux. The fraction of the flux thermalized in the inner disk was fixed at 0.1. In addition,  the electron temperature kT$_e$ and the radius of the Compton illuminated disk r$_{irr}$ were set at 100~keV and 1.1, respectively.  These values are typical for this model \citep[e.g.][]{gie09, gie08, gris12}. 

{\bf  BMC Model:}  We also used the Bulk Motion Comptonization model (BMC in XSPEC), described by \citet{tit97}.  \citet{glio11} have suggested that this model is useful for ULX-like black-hole accretion scenarios.  The best fit is obtained for a fraction of Comptonized flux of 0.12\%.

{\bf SIRF Model:}  Finally we tried the Self-IrRadiated Funnel model (SIRF in XSPEC), which describes the emission from a ``supecritical funnel'' \citep{abol09}. This is created in objects similar to SS~433, by the wind formed at high accretion rates. \citet{abol09} obtained a reasonable fit for XMM-Newton data of NGC 6946 X-1. The funnel opening angle was fixed at 20 degrees, the inclination angle was set to zero, and the accretion rate fixed at an Eddington ratio of 10$^4$. These model parameters were set based on the best-fit results to the X-ray data.  We obtained similar results for inner temperature and radius as \citet{abol09}, but unfortunately could not find an acceptable fit using this model as Table~\ref{table2} shows.  As a result, this model has been disregarded as a viable model to fit the NGC 6946 X-1 SED in the discussion to follow.

The results from the X-ray fits are presented in Table~\ref{table2}, and the models are plotted in Figure~\ref{sed}. The first two models, DISKIR and BMC both provide equally good fits to the X-ray data and will be used to construct the input Spectral Energy Distribution (SED) for the CLOUDY modeling. However, as seen in Figure~\ref{sed}, they predict very different UV and optical emission, and therefore should yield very different results in the photoionization simulations.  The physical difference between these two models is that the BMC model requires a bright optical companion to fit the optical and UV data. As seen in Figure~\ref{sed}, the DISKIR model fits the optical and UV data equally well with a bright (supergiant) companion (dotted red line) or an irradiated disk with a less luminous companion (solid red line) .  

The model for the O9~I supergiant shown as the solid line in blue in Figure~\ref{sed} is a TLUSTY \citep{hub95} O star model with effective temperature 31000~K, log~g~$=$~3.5, and a luminosity of 3.16~$\times$~10$^{39}$ erg~s$^{-1}$.  This fits the optical HST data (round black circles in Figure~\ref{sed}) very well.  We add this star to both DISKIR and BMC models (dotted lines), but we remind our reader that the irradiated disk alone (solid red line) fits the optical-UV data equally well.
Finally, we find that the SIRF model peaks just below the [O~IV] edge, and is faint in the optical-UV, and therefore does not provide a good fit for this data as described above.

\subsection{CLOUDY Photoionization Modeling}

We use DISKIR and BMC SED models shown in Figure~\ref{sed} as input spectra for CLOUDY.  The bolometric luminosities that result from these models are 8.02 and 2.51~$\times$~10$^{40}$ erg~s$^{-1}$, respectively. We assumed a Solar metallicity as per \citet{gar10}.  Previous optical studies of NGC 6946 X-1 \citep{dunne00, blair01, abol08} indicate that the ionized gas from the nebula was swept inside a thin shell ($\sim$~1~pc in width) around the ULX.

\begin{figure}
\epsscale{1.0}
\plottwo{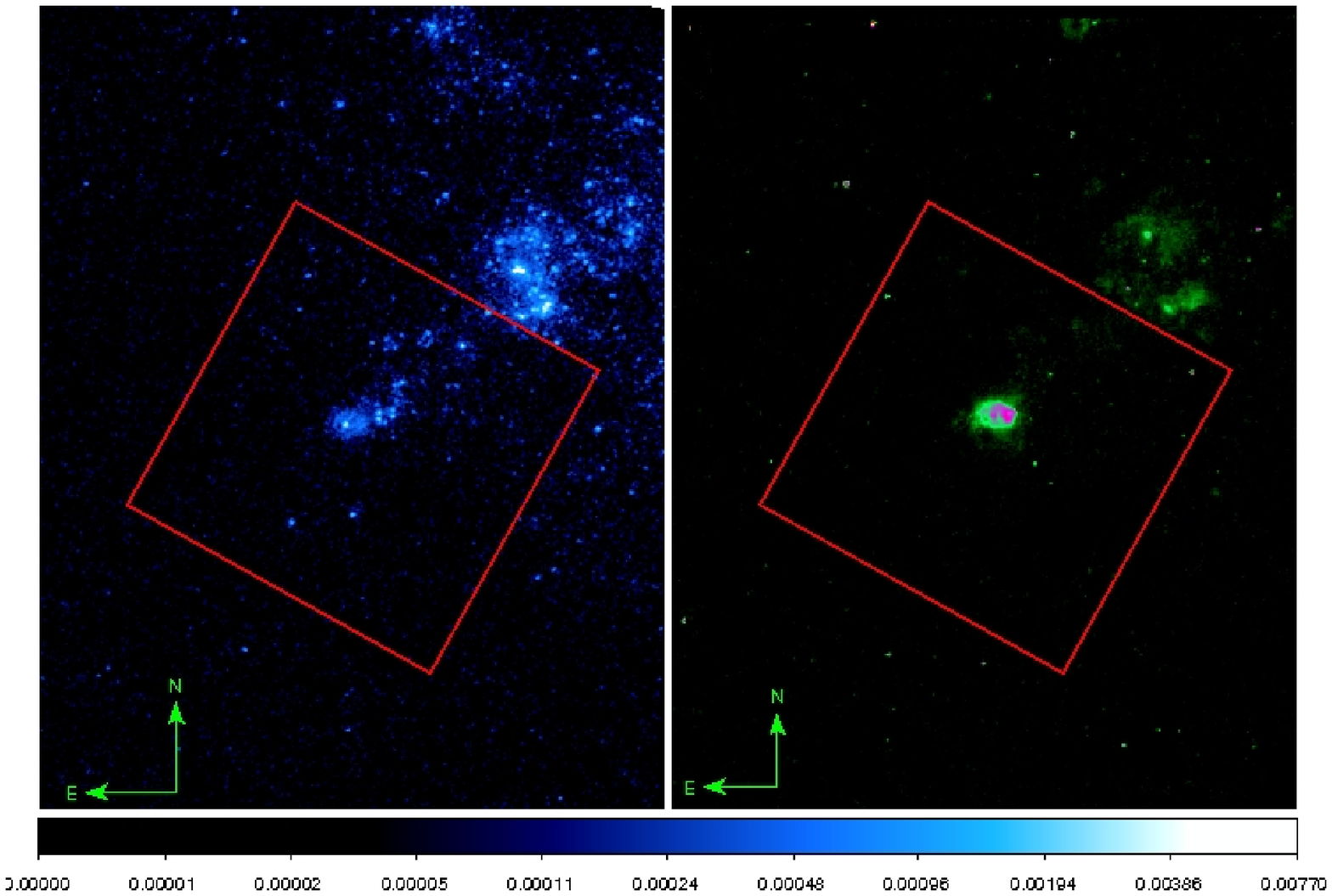}{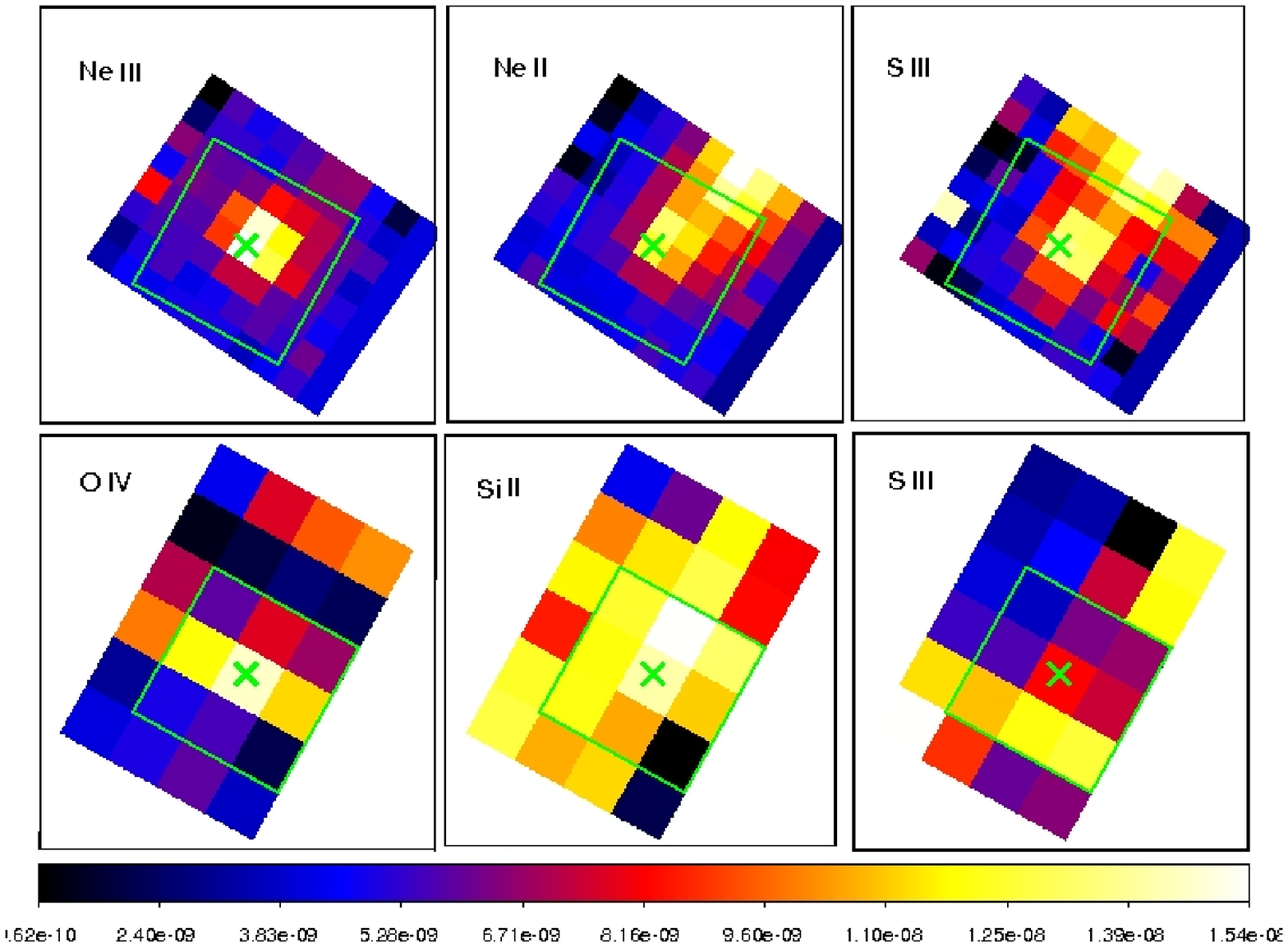}
\caption{
{\it HST} ACS images of the MF 16 nebula in NGC~6946 (left), 
and the {\it Spitzer} spectral maps of the infrared detected lines (right).
The {\it HST} images are UV F140PL filter on the left and H$\alpha$ on the right.
The aperture used to extract IRS spectra is overplotted on all images.
An OB association of bright stars are seen on the UV image just outside the aperture box.
The low-ionization lines are clearly contaminated by these stars.
The associated HII region is seen on the H$\alpha$ image. 
The spectral maps are: [Ne III] 15.56~$\mu$m, [Ne II] 12.81~$\mu$m, [S III] 18.71~$\mu$m, [O IV] 25.89~$\mu$m, 
[S III] 33.48~$\mu$m and [Si II] 34.82~$\mu$m.
The ULX UV counterpart identified by \citet{kaa10} has the appearance of a bright star in the {\it HST} image,
close to the center of the MF~16 nebula, and its position is shown with a green X on the spectral maps. 
} \label{maps}
\end{figure}

\begin{figure}
\epsscale{0.7}
\plotone{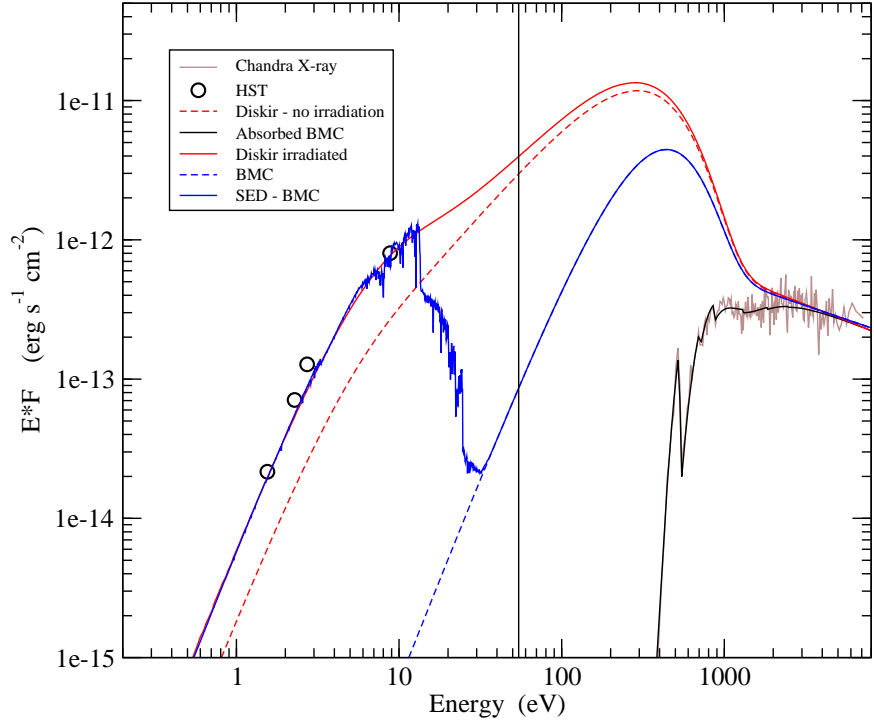}
\caption{
This figure shows the SED constructed in Section~3.
The Hubble optical and UV data are quoted from \citet{kaa10}.
The base model selected as the best fit from the Cloudy simulation (blue) is based on the BMC
model fit to the X-ray data and a O9~I supergiant TLUSTY model star.
We also show a model based on the DISKIR model (continuous red line). 
This includes irradiation of the outer disk replacing the companion 
(the disk without irradiation is shown with a dashed red line)
The [O~IV] edge at 54.93~eV is shown as a vertical line.
} \label{sed}
\end{figure}

Estimates based on line ratios of [SII] suggest that the density of the gas in the shell is quite high (400$-$600 cm$^{-3}$) \citep{blair01, abol08}.  If we use the ratio of the [S~III] IR lines to estimate the density \citep{rachel07}, we obtain densities lower than 150 cm$^{-3}$ for temperatures in the range 10,000$-$20,000~K \citep{abol08}.  However, Figure~\ref{maps} shows that the [S~III] lines are contaminated by the nearby star-forming region and therefore we prefer the higher value obtained from previous optical line ratios estimates, since these were obtained from higher resolution maps.  Thus we assume a spherical shell geometry, a density of 400 cm$^{-3}$, and a filling factor of 0.2 \citep{blair01}.

\begin{figure}
\epsscale{1.0}
\plotone{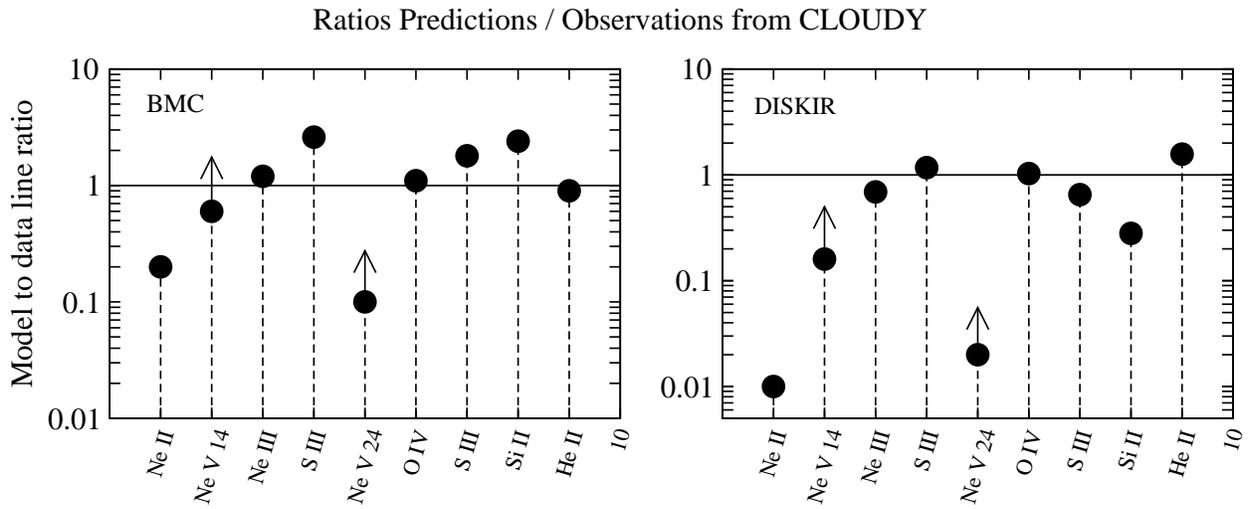}
\caption{
Line predictions from CLOUDY simulations.
We show the results for both input models, BMC and DISKIR, as model-to-data flux ratios.
The perfect match (ratio = 1) is shown as a horizontal line. 
Non-detections are marked with an arrow pointing upward.
} \label{predictions}
\end{figure}

The inner and outer radii of the shell were varied to obtain the best fit for the higher ionization lines [O~IV] and [Ne~III], since these IR-lines are less contaminated by surrounding star-forming regions as illustrated in Figure~\ref{maps}.  We also use the previously measured He~II line flux (1.3 $\times$ 10$^{37}$  erg~s$^{-1}$) as a constraint \citep{wong10}, since this line has an ionization potential similar to [O~IV] and [Ne~III] and likely emanates from a similar region to these IR-lines.  The predicted lines from the CLOUDY simulations are shown in Table~\ref{cloudy} and Figure~\ref{predictions}.

Previous optical observations have shown that radius of the shell (or shells)
is 7$-$10~pc and the thickness is of the order of $\sim$1~pc \citep{blair01}. The DISKIR model is consistent with a shell of radius 20~pc and is thin compared to the findings of \citet{blair01} by a factor of $\sim$10. The BMC model predicts radial dimensions that are closer to those values observed by \citet{blair01}, but the shell predicted is smaller by a factor of $\sim$1.5$-$2. A larger shell than \citet{blair01} is expected for the DISKIR model, because it provides more ionizing photons.  Both models are very good fits to the high ionization IR lines as shown in Figure~\ref{predictions}.  However, the BMC model over-predicts most low-ionization lines, while DISKIR under-predicts them.  

We finally note that both models under-predict the amount of gas compared to the 
Hydrogen column densities obtained from spectral fits. This could be explained if part of the absorption takes place very close to the accretion disk and not near the line-emitting region.

\section{DISCUSSION}

\subsection{The origin of the [O~IV] emission}

\citet{ber10b} discuss in detail possible scenarios for the [O~IV] detected in the Homlberg~II ULX in relation with AGN and star-forming regions.
The conclusions of that paper apply to NGC 6946 X-1 as well. The [O~IV] is clearly associated with the ULX and not with the star-formation nearby (Figure~\ref{maps}). Moreover, the hottest stars alone cannot produce the [O~IV] measured from the MF~16 nebula. 
For more details we again direct our reader to \citet{ber10b} and the references therein. An interesting result of that paper was that hot stars {\it can} have a important effect {\it if} a high energy source is also present. The detailed CLOUDY simulation in \citet{ber10b} showed that an O star can increase the [O~IV] luminosity by a factor of 3. This is also true in our case if the ULX companion is an O-star. We draw a similar conclusion based on the He~II line, which has a very similar ionization potential to [O~IV], and therefore requires a very high energy source.

The results of our CLOUDY modeling cannot easily discriminate between an O-star plus accretion disk and an irradiated disk (Figure~\ref{sed}).  As discussed in Section 3.2, previous optical data suggest that the radius of the shell is 7-10 pc and the thickness is on order 1 pc \citep{blair01}.  Table~\ref{cloudy} shows that the BMC model predicts a radius and thickness that is much closer to the optical findings, than the DISKIR predictions.  Finally, Figure~\ref{predictions} show that the BMC model predictions for He~II and [Ne~III] are very slightly better than the DISKIR model predictions.  For all of these reasons, we prefer the BMC model (the O-star plus accretion disk model) to the DISKIR model (an irradiated disk) to describe NGC~6946 X-1.

\begin{figure}
\epsscale{0.7}
\plotone{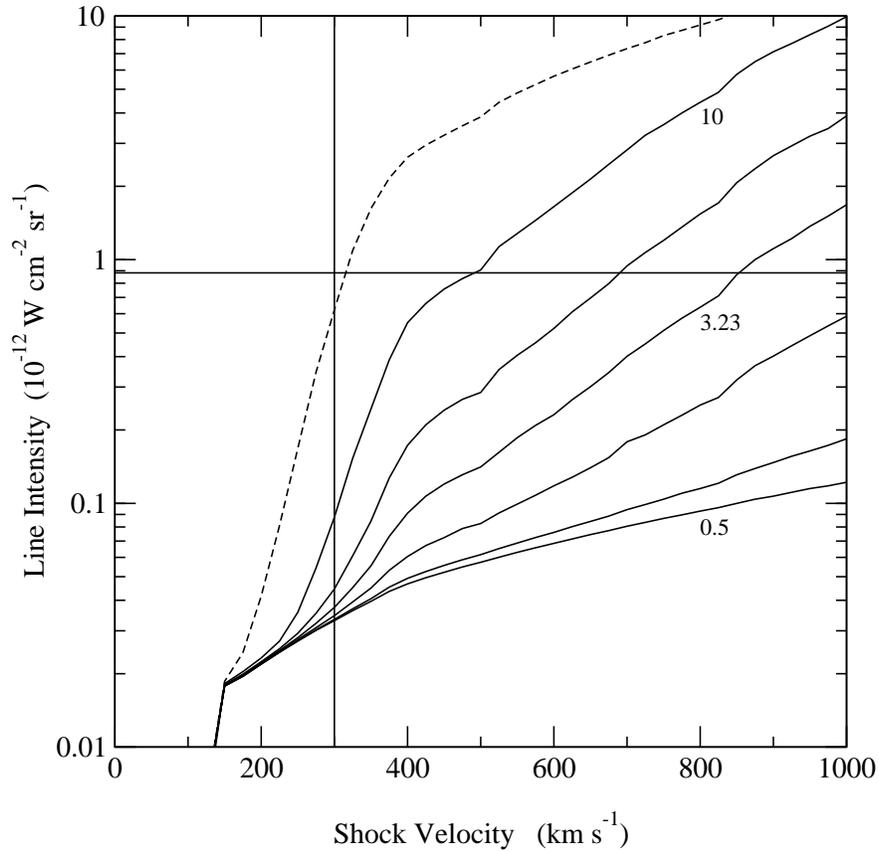}
\caption{
[O~IV] line intensity shocks predictions using Mappings~III.
The shocks plus precursor models vary little with the magnetic parameter, 
and we only show the model with a nominal equipartition value (3.23~$\mu$G~cm$^{3/2}$, the dashed line).
The horizontal line correspond to the brightest pixel in the [O~IV] line map in Fig.~\ref{maps}b.
The vertical line marks the higher limit of 300~km~s$^{-1}$ for shock velocities measured by \citet{dunne00}.
} \label{shocks}
\end{figure}

\citet{ber10b} explored shocks as a possible mechanism for the lines for both NGC~6946 X-1 and Holmberg~II ULXs. They found that the He~II line is probably produced by shocks in the case of another famous nebula-surrounded ULX in Holmberg~IX.
However for NGC 6946 X-1 and Holmberg~II ULXs the He~II emission is consistent with photoionization and not with shocks.  To further confirm that the [O~IV] alone is not dominated by shocks, we reproduced Figure~10a from \citet{ber10b} for NGC~6946 X-1. We used the brightest pixel in the [O~IV] spectral map to estimate a peak emission for this line. We note that this is in fact a lower limit for the estimate, since it represents the average flux over the large LH pixel.
As shown in Figure~\ref{shocks}, the shock velocity required to produce the peak [O~IV] in NGC~6946 X-1 is in excess of 300 km s$^{-1}$.  \citet{dunne00} find shock velocities from H$\alpha$ are at most 270 km s$^{-1}$.  This suggests that the shocks needed to produce the high-ionization lines [O~IV] and He~II are simply not present, based on the H$\alpha$ data.   

The ultimate source of X-ray and high-ionization emission in ULXs has been widely debated.  Indeed, multiple mechanisms have been proposed to explain the high-energy emission, including beamed stellar-mass black holes, super-Eddington stellar mass black holes, and finally, intermediate mass black holes (IMBHs).  Evidence for one or more of these scenarios has been found for a variety of ULXs.  While our observations and modeling of the ULXs in Holmberg~II and the M16 nebulae strongly suggest these objects are not supernova remnants or beamed stellar-mass black holes, this analysis cannot distinguish between a super-Eddington, stellar-mass black hole or an IMBH as the source of the X-ray, optical, and high ionization emission. This is due in part because the X-ray continuum is fit equally well with models describing either scenario. We note on the other hand, that the infamous SS~433 in our Galaxy, is a known beamed, super-Eddington, stellar-mass black hole with a clear host nebula.  All evidence provided in the literature points to shocks or winds/outflows as the primary mechanism(s) for ionization in this object and the nebular region immediately surrounding the ULX \citep[e.g.][]{abol10, beg80, beg06}.  Because SS~433 is the only super-Eddington stellar mass black hole with a nebula definitively known to date, this is the only such source we have to compare with NGC~6946 X-1. Our analysis suggests that unlike SS~433, NGC~6946 X-1 has high~ionization lines \citep[specifically the Oxygen line here and the He~II from][that are most likely {\it not} produced by shocks]{abol08}.
In this singular way the famous SS~433 super-Eddington stellar mass black hole is very different from NGC~6946 X-1.  

The deficiency of super-Eddington sources in nebular environments precludes a detailed comparison of NGC~6946 X-1 with such objects.  However, multiple active black holes in dense environments that produce high-ionization lines via photoionization are known to exist and are well documented in the literature.  These sources, while not IMBHs, may prove an excellent comparative resource.  For this reason, in the subsequent analysis, we focus attention on this comparison.

\subsection{Diagnostic line ratios}

\begin{figure}
\epsscale{1.0}
\plottwo{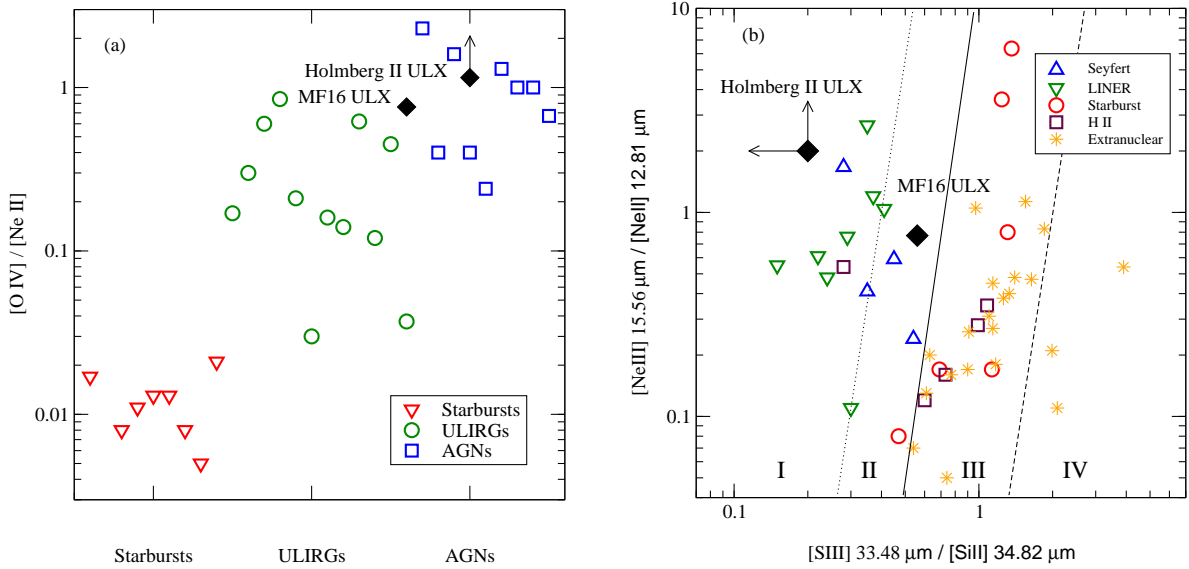}{f6b.eps}
\caption{
Ratio line diagnostics comparing the ULXs NGC 6946 X-1 and Holmberg~II
with other extragalactic ionizing sources: AGN, starbursts, etc.
Both plots suggest the ULXs are very similar to AGNs and LINERs.
a) Diagnostic that uses the [O~IV] line normalized to the [Ne~II] line.
The AGN, starbursts and ULIRG data is taken from \citet{gen98}. 
Many of the starbursts and ULIRGs are actually upper limits \citep[see Fig.~3 of][]{gen98}.
For Holmberg~II ULX we show the lower limit because [Ne~II] was not detected.
b) Diagnostic plot that uses the [S~III] and [Si~II] lines to distinguish between different sources.  
We plot all the data presented by \citet{dale06} (their Fig.~5), which includes: Seyferts, LINERs, starburst galaxies, 
H~II regions and extranuclear star-forming regions.
For Holmberg~II ULX we show the lower and upper limits because [Ne~II] and [S~III], respectively, were not detected.
} \label{ratios}
\end{figure}

The infrared [O~IV] line (together with [Ne~V]) is a signature of high ionization and excitation usually associated with actively accreting black holes \citep[e.g.][]{lutz98}. 
This line is often used to disentangle the starburst emission in composite type galaxies, and is a good indicator of the AGN power \citep{gen98,sturm02,sat04,smith04,sat07, mel08}.  \citet{gen98} and \citet{sturm02} have used the ratio of [O~IV]/[Ne~II] to distinguish between starbursts and AGN dominated galaxies in the mid-IR. In Figure~\ref{ratios}a we compare the NGC 6946 X-1 mid-IR line ratios with data published by \citet{gen98}. The comparison clearly indicates high excitation for the ULX, similar to those seen in AGN. The Holmberg~II ULX from \citet{ber10a} is also plotted here, though the [Ne~II] flux for this ULX is an upper-limit.  Both sources appear as AGN-like when this diagnostic is used.  

Other infrared line diagnostics have been proposed by \citet{dale06, dale08} to disentangle AGN-type ionizing sources from star formation, using the ratios [S~III] 33.48~$\mu$m / [Si~II] 34.82~$\mu$m and [Ne~III] / [Ne~II]. Figure~\ref{ratios}b shows the four regions defined by \citet{dale06} from their Figure~5, and their published data for different sources in the SINGS sample. NGC 6946 X-1 is located well within region II, together with most Seyfert-like nuclei.  Holmberg~II, is also plotted here from \citet{ber10a}, and shows ratios similar to low-luminosity nuclei, but we note that both [S~III] and [Ne~II] are upper-limits for this object.  We also note that the [S~III], [Si~II], and [Ne~II] are heavily contaminated by star formation as is clear in the spectral maps in Figure~\ref{maps}.  However, we also note that the AGN and star forming regions presented by \citet{dale06, dale08} are subject to the same contamination biases and aperture effects as the ULXs.  We therefore still find it useful to compare the two ULXs with these sources.    

The fact that the \citet{dale06, dale08} ratios for NGC~6946 X-1  are Seyfert-like and that the [O~IV]/[Ne~II]  ratio is quite high, suggests that NGC 6946 X-1 is very gas rich compared to most ULXs and has a very strong UV source producing the ionizing radiation.  We note that \citet{vaz07} found similar results in their Spitzer IRS study of the NGC~4485/90 ULXs. However they did not find clear evidence for [O~IV] in their sample of objects.  In examining this data further we find slightly discrepant ratios from \citet{vaz07}.  From our analysis, only ULX-2 in their sample actually falls marginally inside region I, the other five ULXs being located in regions II and III.  In communicating with the authors we find that the discrepancy is likely do to differences in reduction pipeline used for processing or the apertures used in either analysis.

Finally we compared the mid-IR line ratios for NGC 6946 X-1 (and Holmberg~II ULX) with the those of other objects presented in the diagnostic diagrams of \citet{weaver10}.  The error bars for the ULXs in this plot are approximately the same size as the data points and represent 25\% absolute calibration error for CUBISM \citep{smith04}.  \citet{weaver10} find that high-ionization AGNs from the Swift Burst Alert Telescope (BAT) sample \citep{tul08,tul10} distinguish themselves from other, less luminous AGN and starburst galaxies in their [Ne~III]/[Ne~II] and [O~IV]/[Ne~II] ratios.  A replication of the \citet{weaver10} plot is show in Figure~\ref{weaver}.  These high-ionization AGN have ratios very close to or over 1.  In  Figure~\ref{weaver} we over-plotted the NGC~6946 X-1 as well as the Holmberg~II ULX $-$ the only two ULXs known to produce prominent [O~IV] emission.  Interestingly, NGC~6946 X-1 occupies the overlapping region between the high ionization BAT AGN and their lower-luminosity counterparts.  The results for NGC~6946 X-1 are consistent with photoionization which is similar to that seen in LINER or Seyfert-like nuclei rather than star forming in nature.  

All three diagnostic diagrams suggest that the [O~IV] emission in Holmberg~II ULX and in NGC 6946 X-1 is dominated by the ULX and that the surrounding star forming regions provide little contamination to this line. Both ULXs have been shown to be candidate intermediate mass black holes by other authors \citep{leh05, rao10}. While these findings cannot definitively prove that NGC~6946 X-1 is either an IMBH or a super-Eddington accretor, it does suggest that the environment may be very similar for this source and AGN-like objects.  We finally mention that the correlation of NGC~6946 X-1 with ultra-luminous infrared galaxies (ULIRGs) in Figure~\ref{ratios}a and low ionization nuclear emission line regions (LINERs) in Figure~\ref{weaver}, may lend some credence the predictions of \citet{McK10, McK11} who suggest that many LINERs might host ULX-like IMBHs. Indeed many ULIRGs are LINERs \citep{sat07, rachel07, car99}.

\subsection{Black Hole Mass Estimates}

Whether NGC~6946 X-1 is a super-Eddington source or an IMBH cannot be definitely determined from the observed and modeled data presented in this paper. However as discussed in Section 4.2, the ratios from NGC~6946 X-1 look very similar to the ratios of well-known AGN.  We therefore find it useful to estimate the possible black hole mass of this ULX, based on AGN criteria with the caveat that the criteria assumes NGC~6946 X-1 behaves much like an AGN which is not definitely known from the observed and modeled data presented here. The bullets below show four black hole mass estimates derived for NGC~6946 X-1.

\begin{enumerate}
  \item   \citet{glio11, glio09} have proposed a new method to estimate BH masses using the BMC model. This method uses the photon index of the Comptonized part of the spectrum and the normalization to scale from well-known stellar-mass BHs with well-defined spectral transitions.  Using a scaling method defined by \citet{sha09} that is based on a detailed study of 8 different Galactic black hole systems with 17 spectral transitions, \citet{glio09, glio11} extends these scaling techniques to extra-galactic black holes sources. Indeed one of these sources is a newly discovered ULX in NGC~3621.  Thus this technique can be used in the case of NGC~6946 X-1 as well, with the assumption that the states of ULXs mimic those of GBHB.  It assumes that BHs of all masses show similar spectral states, at least with respect to Comptonization. If we scale the results from our BMC fits to XTEJ1550-564 \citep[as done by][in Figure~1]{glio11} and use M$_{\odot}=$9.1 \& d$=$4.4~kpc from  \citet{oro11}, and $\theta=$72 from \cite{glio11}, we obtain a BH mass of $\approx$1,800~M$_{\odot}$.

  \item Both DISKIR and BMC X-ray models predict a cool disk. Assuming a BH accreting at the Eddington limit, the mass of the BH can be estimated at 1$-$1.5$\times$10$^{4}$~M$_{\odot}$. However, this estimate assumes that the ULX is in a typical high state.  If this is not the case, the inner disk temperature cannot be used reliably to estimate the mass of the BH \citep{wil06}. 

  \item Another estimate can be made just based on the bolometric luminosity calculated from our models. Using the values listed in Table~\ref{table2}, and assuming again accretion at Eddington limit, we obtain masses of 130 and 450~M$_{\odot}$.  

  \item Finally, \citet{das08} found a good correlation between the luminosity of the [O~IV] line and the mass of the central BH in standard AGN. We estimate a BH mass of 2.7$\times$10$^{5}$~M$_{\odot}$ if this correlation is extrapolated to NGC 6946 X-1. 

In conclusion these four methods predict a very wide range of black hole masses for NGC~6946 X-1.  Indeed the estimates predict masses that span three orders of magnitude:  10$^{2}$ to 10$^{5}$M$_{\odot}$, so emphasis on any one estimate is difficult to justify.  However it is interesting to note that all four put this object within the IMBH regime for the given criteria (e.g. accretion at the Eddington rate).  

\end{enumerate}

\section{CONCLUSIONS}

Using the mid-IR {\it Spitzer} spectral maps of NGC~6946 X-1 combined with {\it Chandra} X-ray and {\it HST} optical data from the literature, we model the spectral energy distribution of this unique ULX. We find the following major conclusions from our modeling and observational results:

\begin {itemize}
\item  The SED constructed based on X-ray and optical spectral modeling indicates that NGC 6946 X-1 is well fit with two Comptonization models:  The DISKIR model and the BMC model, both available in XSPEC.  Spectral modeling did not permit a good distinction between an irradiated disk or a bright companion plus accretion disk as source of the optical emission.  However, the nebula parameters (namely radius and thickness) predicted by CLOUDY using the BMC model better match the findings from optical observations \citep{blair01} and the lines are slightly better matched with the BMC model in this case. We therefore prefer the BMC model to the DISKIR model.
\item The detected [O~IV] is consistent with photoionization by the strong X-ray emission from the ULX, and not with shocks or nearby star-formation. This suggests that the X-ray emission is not strongly beamed. The data cannot however discriminate between super-Eddington accretion and the IMBH scenario. 
\item We compared our results with known super-Eddington sources (i.e. SS~433) as well as standard AGN. We find that the emission line properties of NGC~6946 X-1 are not well matched to the SS~433 super-Eddington source, since the lines from this source have been shown to be produced primarily by shocks.  We also compared NGC~6946 X-1 line ratios with multiple AGN diagnostics.  The line ratios from NGC 6946 X-1 consistently show emission very similar to most low-luminosity AGN.  We plotted diagnostic ratios from \citet{gen98} and \citet{sturm02}, \citet{dale06}, and \citet{weaver10}.  All plots show this ULX falling safely within the AGN regime rather than star forming regions.  This suggests that NGC~6946 X-1 may have photoionization and nebula properties that are similar to those seen in AGN.  
\item  A variety of approximations were used to estimate the black hole mass from the observed properties of this ULX and resulted in a predicted black hole mass range spanning three orders of magnitude.  Assuming this ULX is accreting at the Eddington limit we find that NGC 6946 X-1 has a black hole mass between 10$^{2}$ to 10$^{5}$ M$_{\odot}$.  Our calculations suggest that NGC~6946 X-1 could have a black hole mass within the IMBH regime, with the caveat that the criteria for estimating black hole mass assumes NGC~6946 X-1 behaves much like an AGN which is not definitely known from the observed and modeled data presented here.

\end{itemize}

\acknowledgments

\begin{figure}
\epsscale{0.7}
\plotone{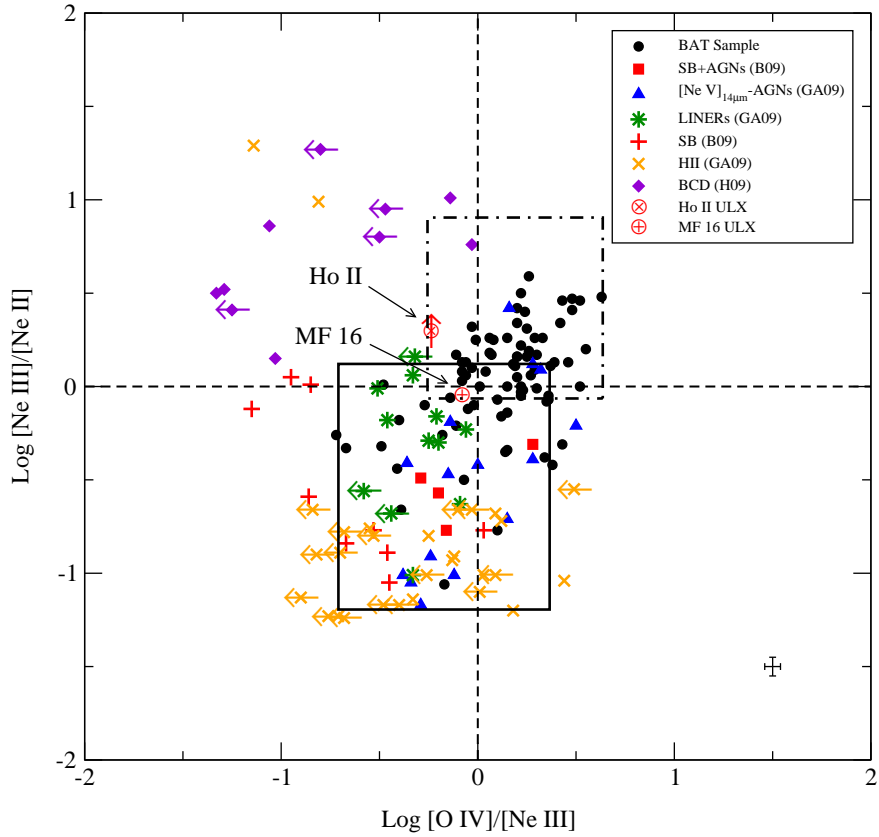}
\caption{
Line ratio diagnostics from \citet{weaver10}.
NGC 6946 X-1 line ratios clearly fit within the overlapping high-ionization and low-ionization AGN regime. 
This is only the second ULX to show such line ratios (see also Holmberg~II).
} \label{weaver}
\end{figure}

We are extremely grateful to Marcio Mel{\'e}ndez for all of his help interpreting and plotting Figure~\ref{weaver}. We are also very grateful to the referee whose comments have greatly improved the quality and clarity of this paper. 
This research has made use of the NASA/IPAC Extragalactic Database (NED) 
which is operated by the Jet Propulsion Laboratory, California Institute of Technology, under contract with the National Aeronautics and Space Administration. This work is based on observations made with the Spitzer Space Telescope, which is operated by the Jet Propulsion Laboratory, California Institute of Technology under a contract with NASA. SMART was developed by the IRS Team at Cornell University and is available through the Spitzer Science Center at Caltech.

\clearpage


\begin{deluxetable}{c|ccc}
\tablecolumns{3}  
\tablewidth{0pt} 
\tabletypesize{\tiny}
\setlength{\tabcolsep}{0.05in}
\tablenum{1} 
\tablecaption{Measured infrared lines\label{table1}}   
\tablehead{
\colhead{Line} & \colhead{Flux} & \colhead{S/N Ratio} & \colhead{L} \\
\colhead{(1)} & \colhead{(2)} & \colhead{(3)} & \colhead{(4)} \\
}
\startdata

$[$Ne II$]$  12.81~$\mu$m  &  2.5$\pm$0.3   &  13.5  &  7.9$\pm$0.9   \\

$[$Ne V$]$   14.32~$\mu$m  &  $<$0.5	    &  ...   &  $<$1.5        \\
$[$Ne III$]$ 15.56~$\mu$m  &  2.3$\pm$0.2   &  18.9  &  7.1$\pm$0.7   \\
$[$S III$]$  18.71~$\mu$m  &  1.3$\pm$0.2   &  12.5  &  4.1$\pm$0.7   \\
$[$Ne V$]$   24.32~$\mu$m  &  $<$2.1	    &  ...   &  $<$6.7        \\
$[$O IV$]$   25.89~$\mu$m  &  1.9$\pm$0.2   &  12.8  &  6.0$\pm$0.5   \\
$[$Fe II$]$  25.99~$\mu$m  &  $<$1.2	    &  ...   &  $<$3.7        \\
$[$S III$]$  33.48~$\mu$m  &  3.0$\pm$0.2   &  24.3  &  9.4$\pm$0.6   \\
$[$Si II$]$  34.82~$\mu$m  &  2.4$\pm$0.4   &  9.90  &  7.4$\pm$1.4   \\

\enddata

\tablecomments{   
Fluxes are in 10$^{-21}$~W~cm$^{-2}$, luminosities (L) in 10$^{37}$~erg~s$^{-1}$.
If the measurements errors are smaller than the absolute calibration accuracy of 25\%, 
the latter were used. For nondetections we show 3$\sigma$ upper limits.
}
\end{deluxetable}


\begin{deluxetable}{l|ccccc}
\tablecolumns{5}  
\tablewidth{0pt} 
\tabletypesize{\tiny}
\setlength{\tabcolsep}{0.05in}
\tablenum{2} 
\tablecaption{X-ray model fits\label{table2}}   
\tablehead{
\colhead{Model} & \colhead{N$_H$} & \colhead{kT$_{in}$} & \colhead{$\Gamma$/R$_{in}$} & \colhead{$\Delta\chi^2$/dof} & \colhead{L} \\
\colhead{(1)} & \colhead{(2)} & \colhead{(3)} & \colhead{(4)} & \colhead{(5)} & \colhead{(6)} \\
\colhead{} & \colhead{(10$^{21}$ cm$^{-3}$)} & \colhead{(keV)} & \colhead{} & \colhead{} & \colhead{(10$^{40}$ erg s$^{-1}$)} \\
}
\startdata

DISKIR	   &  4.26$\pm$0.11           &	0.118$\pm$0.003         &  2.44$\pm$0.08	   &  1.16/183	&  7.01 \\
BMC	   &  3.32$^{+0.69}_{-0.39}$  &	0.110$\pm$0.008         &  2.40$^{+0.08}_{-0.10}$  &  1.17/182	&  2.12 \\
SIRF       &  1.63$^{+0.12}_{-0.08}$  &	1.45$^{+0.32}_{-0.11}$  &  7.89$^{+10.9}_{-3.8}$   &  1.85/182	&  1.39 \\

\enddata

\tablecomments{  
(1): X-ray model, as described in Section~3.1.
(2): Intrinsic hydrogen column density. 
The Galactic column density from \citet{ber08} (2.02$\times$10$^{21}$ cm$^{-2}$) was added separately. 
(3): Inner disk temperature.
(4): Photon index for DISKIR and BMC models and the inner radius in 10$^{-4}$ ``spherisation radius'' units 
for the SIRF model \citep[see][]{abol09}. 
(5): Reduced $\chi^2$ values for the fit and the number of degrees of freedom. 
(6): Unabsorbed (intrinsic) luminosities between 0.1 and 10~keV.
}
\end{deluxetable}

\begin{deluxetable}{l|cccccccccccc}
\tablecolumns{12}  
\tablewidth{0pt} 
\tabletypesize{\tiny}
\setlength{\tabcolsep}{0.05in}
\tablenum{3} 
\tablecaption{CLOUDY modeling results\label{cloudy}}   
\tablehead{
\colhead{Model} & \colhead{Radius} & \colhead{Depth}
& \colhead{He~II}
& \colhead{$[$Ne II$]$} & \colhead{$[$Ne V$]$} & \colhead{$[$Ne III$]$} & \colhead{$[$S III$]$}
& \colhead{$[$Ne V$]$} & \colhead{$[$O IV$]$} & \colhead{$[$Fe II$]$} & \colhead{$[$S III$]$}
& \colhead{$[$Si II$]$} \\
\colhead{} & \colhead{} & \colhead{}
& \colhead{$\lambda$4686} 
& \colhead{12.81} & \colhead{14.32} & \colhead{15.56} & \colhead{18.71} 
& \colhead{24.32} & \colhead{25.89} & \colhead{25.99} & \colhead{33.48} 
& \colhead{34.82} \\
\colhead{(1)} & \colhead{(2)} & \colhead{(3)} & \colhead{(4)} & \colhead{(5)} 
& \colhead{(6)} 
& \colhead{(7)} & \colhead{(8)} & \colhead{(9)} & \colhead{(10)} 
& \colhead{(11)} & \colhead{(12)} & \colhead{(13)} \\
\colhead{} & \colhead{(pc)} & \colhead{(pc)}
& \colhead{}
& \colhead{} & \colhead{} & \colhead{} & \colhead{} 
& \colhead{} & \colhead{} & \colhead{} & \colhead{} 
& \colhead{} \\
}
\startdata

BMC    & 5.1  & 2.6  &  37.14 & 37.16  &  36.97  &  37.93  &  38.02  &  37.04  &  37.82  &  37.86  &  38.24  &  38.25 \\
DISKIR & 20.7 & 0.06 &  37.40 & 35.98  &  36.38  &  37.69  &  37.68  &  36.45  &  37.79  &  36.57  &  37.79  &  37.29 \\

\enddata
\tablecomments{  
(1): SED model, as described in Section~3.1.
(2): Inner shell radius.
(3): The thickness of the shell.
Line luminosities are in units of log erg s$^{-1}$.
}
\end{deluxetable}


\begin{thebibliography}{}

\bibitem[Abolmasov et al.(2007)]{abol07} Abolmasov, P., Fabrika, S., Sholukhova, O., \& Afanasiev, V.\ 2007, Astrophysical Bulletin, 62, 36 
\bibitem[Abolmasov et al.(2008)]{abol08} Abolmasov, P., Fabrika, S., Sholukhova, O., \& Kotani, T.\ 2008, arXiv:0809.0409
\bibitem[Abolmasov et al.(2009)]{abol09} Abolmasov, P., Karpov, S., \& Kotani, T.\ 2009, \pasj, 61, 213 
\bibitem[Abolmasov et al.(2010)]{abol10} Abolmasov, P., Maryeva, O., \& Burenkov, A.~N.\ 2010, Astronomische Nachrichten, 331, 412 
\bibitem[Abolmasov(2011)]{abol11} Abolmasov, P.\ 2011, \na, 16, 138 
\bibitem[Begelman et al.(1980)]{beg80} Begelman, M.~C., Hatchett, S.~P., McKee, C.~F., Sarazin, C.~L., 
\& Arons, J.\ 1980, \apj, 238, 722 
\bibitem[Begelman et al.(2006)]{beg06} Begelman, M.~C., King, A.~R., \& Pringle, J.~E.\ 2006, \mnras, 370, 399 
\bibitem[Berghea et al.(2008)]{ber08} Berghea, C.~T., Weaver, K.~A., Colbert, E.~J.~M., \& Roberts, T.~P.\ 2008, \apj, 687, 471 
\bibitem[Berghea et al.(2010a)]{ber10a} Berghea, C.~T., Dudik, R.~P., Weaver, K.~A., \& Kallman, T.~R.\ 2010, \apj, 708, 354 
\bibitem[Berghea et al.(2010b)]{ber10b} Berghea, C.~T., Dudik, R.~P., Weaver, K.~A., \& Kallman, T.~R.\ 2010, \apj, 708, 364 
\bibitem[Blair et al.(2001)]{blair01} Blair, W.~P., Fesen, R.~A., \& Schlegel, E.~M.\ 2001, \aj, 121, 1497 
\bibitem[Carrillo et al.(1999)]{car99} Carrillo, R., Masegosa, J., Dultzin-Hacyan, D., \& Ordo{\~n}ez, R.\ 1999, \rmxaa, 35, 187 
\bibitem[Colbert \& Mushotzky(1999)]{col99} Colbert, E.~J.~M., \& Mushotzky, R.~F. 1999, \apj, 519, 89
\bibitem[Cseh et al.(2012)]{cseh12} Cseh, D., Corbel, S., Kaaret, P., et al.\ 2012, arXiv:1201.4473 
\bibitem[Dale et al.(2006)]{dale06} Dale, D.~A., et al.\ 2006, \apj, 646, 161 
\bibitem[Dale et al.(2008)]{dale08} Dale, D.~A., et al.\ 2008, arXiv:0811.4190 
\bibitem[Dasyra et al.(2008)]{das08} Dasyra, K.~M., et al.\ 2008, \apjl, 674, L9 
\bibitem[Done \& Kubota(2006)]{done06} Done, C., \& Kubota, A. 2006, \mnras, 371, 1216
\bibitem[Draine(1989)]{draine} Draine, B.~T.\ 1989, Infrared Spectroscopy in Astronomy, 290, 93 
\bibitem[Dubner et al.(1998)]{dub98} Dubner, G.~M., Holdaway, M., Goss, W.~M., \& Mirabel, I.~F.\ 1998, \aj, 116, 1842 
\bibitem[Dudik et al.(2007)]{rachel07} Dudik, R.~P., Weingartner, J.~C., Satyapal, S., Fischer, J., Dudley, C.~C., \& O'Halloran, B.\ 2007, \apj, 664, 71
\bibitem[Dunne et al.(2000)]{dunne00} Dunne, B.~C., Gruendl, R.~A., \& Chu, Y.-H.\ 2000, \aj, 119, 1172 
\bibitem[Fabrika(2004)]{fab04} Fabrika, S.\ 2004, Astrophysics and Space Physics Reviews, 12, 1 
\bibitem[Fender \& Belloni(2004)]{fen04} Fender, R., \& Belloni, T.\ 2004, \araa, 42, 317 
\bibitem[Filippeko \& Ho(2003)]{fil03} Filippenko, A.~V., \& Ho, L.~C.\ 2003, \apjl, 588, L13 
\bibitem[Fridriksson et al.(2008)]{frid08} Fridriksson, J.~K., 
Homan, J., Lewin, W.~H.~G., Kong, A.~K.~H., 
\& Pooley, D.\ 2008, \apjs, 177, 465 
\bibitem[Garc{\'{\i}}a-Benito et al.(2010)]{gar10} Garc{\'{\i}}a-Benito, R., et al.\ 2010, \mnras, 408, 2234 
\bibitem[Genzel et al.(1998)]{gen98} Genzel et al. 1998, \apj, 498, 579
\bibitem[Gierli{\'n}ski et al.(2008)]{gie08} Gierli{\'n}ski, M., Done, C., \& Page, K.\ 2008, \mnras, 388, 753
\bibitem[Gierli{\'n}ski et al.(2009)]{gie09} Gierli{\'n}ski, M., Done, C., \& Page, K.\ 2009, \mnras, 392, 1106 
\bibitem[Gliozzi et al.(2009)]{glio09} Gliozzi, M., Satyapal, S., Eracleous, M., Titarchuk, L., \& Cheung, C.~C.\ 2009, \apj, 700, 1759 
\bibitem[Gliozzi et al.(2011)]{glio11} Gliozzi, M., Titarchuk, L., Satyapal, S., Price, D., \& Jang, I.\ 2011, \apj, 735, 16 
\bibitem[Gris{\'e} et al.(2012)]{gris12} Gris{\'e}, F., Kaaret, P., Corbel, S., et al.\ 2012, \apj, 745, 123 
\bibitem[Higdon et al.(2004)]{hig04} Higdon, S.~J.~U., et al.\ 2004, \pasp, 116, 975 
\bibitem[Hubeny \& Lanz(1995)]{hub95} Hubeny, I., \& Lanz, T.\ 1995, \apj, 439, 875 
\bibitem[Kaaret et al.(2004)]{kaa04} Kaaret, P., Ward, M., \& Zezas, A. 2004, \mnras, 351, L83
\bibitem[Kaaret \& Corbel(2009)]{kaa09} Kaaret, P., \& Corbel, S.\ 2009, \apj, 697, 950 
\bibitem[Kaaret et al.(2010)]{kaa10} Kaaret, P., Feng, H., Wong, D.~S., \& Tao, L.\ 2010, \apjl, 714, L167 
\bibitem[Kennicutt et al.(2003)]{ken03} Kennicutt R.C. et al. 2003, PASP, 115, 928
\bibitem[King(2008)]{king08} King, A.~R.\ 2008, \mnras, 385, L113
\bibitem[K{\"o}rding et al.(2002)]{kor02} K{\"o}rding, E., Falcke, H., \& Markoff, S.\ 2002, \aap, 382, L13
\bibitem[Kuncic et al.(2007)]{kun07} Kuncic, Z., Soria, R., Hung, C.~K., Freeland, M.~C., \& Bicknell, G.~V.\ 2007, IAU Symposium, 238, 247 
\bibitem[Lehmann et al.(2005)]{leh05} Lehmann, I. et al. 2005, A\&A, 431, 847
\bibitem[Lutz et al.(1998)]{lutz98} Lutz, D., Kunze, D., Spoon, H.~W.~W., \& Thornley, M.~D. 1998, A\&A, 333, L75
\bibitem[Matonick \& Fesen(1997)]{mf97} Matonick, D.~M., \& Fesen, R.~A.\ 1997, \apjs, 112, 49 
\bibitem[McKernan et al.(2010)]{McK10} McKernan, B., Ford,  K.~E.~S., \& Reynolds, C.~S.\ 2010, \mnras, 407, 2399 
\bibitem[McKernan et al.(2011)]{McK11} McKernan, B., Ford, K.~E.~S., Yaqoob, T., \& Winter, L.~M.\ 2011, \mnras, 413, L24 
\bibitem[Mel{\'e}ndez et al.(2008)]{mel08} Mel{\'e}ndez, M., Kraemer, S.~B., Schmitt, H.~R., Crenshaw, D.~M., Deo, R.~P., Mushotzky, R.~F., \& Bruhweiler, F.~C.\ 2008, \apj, 689, 95 
\bibitem[Miller et al.(2004)]{mill04} Miller, J.~M., Fabian, A.~C., \& Miller, M.~C. 2004, \apj, 614, L117
\bibitem[Miller et al.(2005)]{mill05} Miller, N.~A., Mushotzky, R.~F., \& Neff, S.~G. 2005, \apjl, 623, L109
\bibitem[Mitsuda, et al.(1984)]{mit84} Mitsuda, K. et al. 1984, PASJ, 36, 741
\bibitem[Moon et al.(2011)]{moon11} Moon, D.-S., Harrison, F.~A., Cenko, S.~B., \& Shariff, J.~A.\ 2011, \apjl, 731, L32
\bibitem[Orosz et al.(2011)]{oro11} Orosz, J.~A., Steiner, J.~F., McClintock, J.~E., et al.\ 2011, \apj, 730, 75 
\bibitem[Pakull \& Mirioni(2002)]{pak02} Pakull, M., \& Mirioni, L. 2002, in ``New Visions of the X-ray Universe in the XMM-Newton and Chandra Era'', (Noordwijk: ESTEC), (astro-ph/0202488)
\bibitem[Pakull et al.(2010)]{pak10} Pakull, M.~W., Soria, R., \& Motch, C.\ 2010, \nat, 466, 209 
\bibitem[Poutanen et al.(2007)]{pou07} Poutanen, J., Lipunova, G., Fabrika, S., Butkevich, A.~G., \& Abolmasov, P.\ 2007, \mnras, 377, 1187
\bibitem[Rao et al.(2010)]{rao10} Rao, F., Feng, H., \& Kaaret, P.\ 2010, \apj, 722, 620  
\bibitem[Revnivtsev et al.(2002)]{rev02} Revnivtsev, M., Sunyaev, R., Gilfanov, M., \& Churazov, E.\ 2002, \aap, 385, 904 
\bibitem[Roberts \& Colbert(2003)]{rob03} Roberts, T.~P., \& Colbert, E.~J.~M.\ 2003, \mnras, 341, L49 
\bibitem[Satyapal et al.(2004)]{sat04} Satyapal, S., Sambruna, R.~M., \& Dudik, R.~P.\ 2004, \aap, 414, 825 
\bibitem[Satyapal et al.(2007)]{sat07} Satyapal, S., Vega, D., Heckman, T., O'Halloran, B., \& Dudik, R. 2007, \apj, 663, L9
\bibitem[Shaposhnikov \& Titarchuk(2009)]{sha09} Shaposhnikov, N., \& Titarchuk, L.\ 2009, \apj, 699, 453
\bibitem[Smith	et al.(2004)]{smith04} Smith, J.~D.~T.  et al. 2004, \apjs, 154, 199
\bibitem[Smith et al.(2007)]{smith07} Smith, J.~D.~T., et al.\ 2007, \pasp, 119, 1133 
\bibitem[Soria \& Kuncic(2008)]{sor08} Soria, R., \& Kuncic, Z.\ 2008, American Institute of P\bibitem[Shaposhnikov 
\& Titarchuk(2009)]{2009ApJ...699..453S} Shaposhnikov, N., \& Titarchuk, L.\ 2009, \apj, 699, 453hysics Conference Series, 1053, 103
\bibitem[Soria et al.(2010)]{sor10} Soria, R., Pakull, M.~W., Broderick, J.~W., Corbel, S., \& Motch, C.\ 2010, \mnras, 409, 541 
\bibitem[Stephenson \& Sanduleak(1977)]{sanduleak} Stephenson, C.~B., \& Sanduleak, N.\ 1977, \apjs, 33, 459 
\bibitem[Stobbart et al.(2006)]{wil06} Stobbart, A-M., Roberts, T.~P., \& Wilms, J. 2006, \mnras, 368, 397
\bibitem[Sturm et al.(2002)]{sturm02} Sturm, E., Lutz, D., Verma, A., Netzer, H., Sternberg, A., Moorwood, A.~F.~M., Oliva, E., \& Genzel, R. 2002, A\&A, 393, 821
\bibitem[Titarchuk et al.(1997)]{tit97} Titarchuk, L., Mastichiadis, A., \& Kylafis, N.~D.\ 1997, \apj, 487, 834
\bibitem[Tueller et al.(2008)]{tul08} Tueller, J., Mushotzky, R.~F., Barthelmy, S., et al.\ 2008, \apj, 681, 113 
\bibitem[Tueller et al.(2010)]{tul10} Tueller, J., Baumgartner, W.~H., Markwardt, C.~B., et al.\ 2010, \apjs, 186, 378 
\bibitem[van Dyk et al.(1994)]{dyk94} van Dyk, S.~D., Sramek, R.~A., Weiler, K.~W., Hyman, S.~D., \& Virden, R.~E.\ 1994, \apjl, 425, L77 
\bibitem[V{\'a}zquez et al.(2007)]{vaz07} V{\'a}zquez, G.~A., Hornschemeier, A.~E., Colbert, E., Roberts, T.~P., Ward, M.~J., \& Malhotra, S.\ 2007, \apjl, 658, L21 
\bibitem[Weaver et al.(2010)]{weaver10} Weaver, K.~A., Mel{\'e}ndez, M., Mushotzky, R.~F., et al.\ 2010, \apj, 716, 1151
\bibitem[Wong(2010)]{wong10} Wong, D.\ 2010, 38th COSPAR Scientific Assembly, 38, 2461 


\end{thebibliography}
\end{document}